\documentclass[prl,aps,preprint,showpacs,superscriptaddress]{revtex4}
\usepackage{amssymb,amsmath}
\usepackage{graphicx}
\usepackage{color}

\def\ds{\displaystyle}

\begin{document}

\bibliographystyle{apsrev}

\title{Surface waves versus negative refractive index in layered superconductors}

\author{V.A.~Golick}
\affiliation{V.N.~Karazin Kharkov National University, 61077
Kharkov, Ukraine}

\author{D.V.~Kadygrob}
\affiliation{A.Ya.~Usikov Institute for Radiophysics and
Electronics, Ukrainian Academy of Sciences, 61085 Kharkov,
Ukraine}

\author{V.A.~Yampol'skii}
\affiliation{V.N.~Karazin Kharkov National University, 61077
Kharkov, Ukraine} \affiliation{A.Ya.~Usikov Institute for
Radiophysics and Electronics, Ukrainian Academy of Sciences, 61085
Kharkov, Ukraine} \affiliation{Advanced Science Institute, The
Institute of Physical and Chemical Research (RIKEN), Saitama,
351-0198, Japan}

\author{A.L.~Rakhmanov}
\affiliation{Advanced Science Institute, The Institute of Physical
and Chemical Research (RIKEN), Saitama, 351-0198, Japan}
\affiliation{Institute for Theoretical and Applied Electrodynamics
Russian Acad.~Sci., 125412 Moscow, Russia}

\author{ B.A.~Ivanov}
\affiliation{Advanced Science Institute, The Institute of Physical
and Chemical Research (RIKEN), Saitama, 351-0198, Japan}
\affiliation{Institute of Magnetism, Ukrainian Academy of
Sciences, 03142 Kiev, Ukraine} \affiliation{National T. Shevchenko
University of Kiev, 03127 Kiev, Ukraine}

\author{Franco Nori}
\affiliation{Advanced Science Institute, The Institute of Physical
and Chemical Research (RIKEN), Saitama, 351-0198, Japan}
\affiliation{Department of Physics, University of Michigan, Ann
Arbor, MI 48109, USA}

\begin{abstract}

We predict a new branch of surface Josephson plasma waves (SJPWs)
in layered superconductors for frequencies higher than the
Josephson plasma frequency. In this frequency range, the
permittivity tensor components along and transverse to the layers
have different signs, which is usually associated with negative
refraction. However, for these frequencies, the bulk Josephson
plasma waves cannot be matched with the incident and reflected
waves in the vacuum, and, instead of the negative-refractive
properties, abnormal surface modes appear within the frequency
band expected for bulk modes. We also discuss the excitation of
high-frequency SJPWs by means of the attenuated-total-reflection
method.

\end{abstract}

\maketitle

High-$T_c$ layered cuprate superconductors are important
candidates for negative-refractive-index (NRI) metamaterials (see,
e.g., \cite{supra,negref}). Indeed, being uniaxial strongly
anisotropic materials, they provide different signs of the
permittivity tensor components along, $\varepsilon_{ab}$, and
transverse, $\varepsilon_c$, to the layers in a wide frequency
range (see, e.g.,~\cite{anisa1}), providing a possibility of NRI.
These metamaterials are attracting considerable attention because
they have the potential for subwavelength resolution and
aberration-free imaging. Layered superconductors are very
promising metamaterials because they are relatively
straightforward to fabricate (compared to double negative
metamaterials) and do not require negative permeability.

Experiments for the $\mathbf{c}$-axis conductivity in layered
superconductors prove the use of a model in which the
superconducting CuO$_2$ layers are coupled by the intrinsic
Josephson effect through the layers~\cite{Kl-Mu}. Thus, a
Josephson plasma with anisotropic current capability is produced
in layered superconductors. Moreover, the physical mechanisms of
the currents along and across the layers are fundamentally
different. The current along the layers is similar to the current
in bulk superconductors, while the current across the layers is
Josephson-type.

The Josephson current along the $\mathbf{c}$-axis couples with the
electromagnetic field inside the insulating dielectric layers,
forming Josephson plasma waves (JPWs) (see the
review~\cite{Thz-rev-2008-July} and references therein). Thus, the
propagation of electromagnetic waves through the layers is favored
by the layered structure. The study of these waves is very
important because of their Terahertz frequency range, which is
still hardly reachable for electronic and optical devices.

Like in common plasma waves, JPWs propagate with frequencies above
some threshold value (the Josephson plasma frequency
$\omega_{J}$). However, in the frequency range below $\omega_{J}$,
the presence of the sample boundary can produce surface Josephson
plasma waves (SJPWs)~\cite{mdyam1}, which are analog to the
surface plasmon polaritons in metals~\cite{Plasman,Agrmils}. Such
waves can propagate along the vacuum-layered superconductor
interface and damp away from it.

At frequencies $\omega$ higher than the Josephson plasma frequency
$\omega_J$, the normal-to-the-layers components of both the group
velocity and the Poynting vector of propagating JPWs, have signs
opposite to the sign of the normal component of the wave-vector
$\textbf{k}_s$. This corresponds to a NRI. However, the condition
$\omega > \omega_J$ \emph{is not sufficient} for NRI. This NRI
effect can be observed at the vacuum-layered superconductor
boundary \emph{only in a relatively narrow frequency range},
$\omega_J<\omega<\omega_1^{\rm vac} =\omega_J
[\varepsilon/(\varepsilon - 1)]^{1/2}$, where $\varepsilon$ is the
interlayer dielectric constant. A similar limitation exists for
any insulator-layered superconductor boundary, if the dielectric
constant $\varepsilon_{\rm ext}$ of the insulator is less than
$\varepsilon$. The above conditions follow from the dispersion
relation for the JPWs and the natural limitation for the
tangential component $q$ ($q=k\sin\theta < k, \quad k=\omega/c$)
of the wave-vector for a wave incident at an angle $\theta$ from
the vacuum onto the layered superconductor. A simple analysis
shows that the above inequality is compatible with the dispersion
relation for JPWs only for frequencies
$\omega_J<\omega<\omega_1^{\rm vac} $. In other words, any wave
with frequency $\omega>\omega_1^{\rm vac} $ \emph{incident from
the vacuum} cannot propagate further in a layered superconductor.
JPWs with $\omega>\omega_1^{\rm vac} $ can only match
\emph{evanescent} waves in the vacuum. As for frequencies
$\omega_J<\omega<\omega_1^{\rm vac} $, the NRI can be observed,
but only for incident angles $\theta$ higher than some critical
value $\theta_{\rm crit}$.

For frequencies in the interval
\begin{equation}\label{interval}
\omega_1^{\rm vac} <\omega<\omega_2, \quad \omega_2 = \omega_J\gamma,
\end{equation}
we predict the existence of a \emph{new branch of surface waves}
(here $\gamma=\lambda_c/\lambda_{ab} \gg 1$ is the
current-anisotropy parameter, $\lambda_c
=c/\omega_J\varepsilon^{1/2}$ and $\lambda_{ab}$ are the
magnetic-field penetration depths along and across the layers,
respectively). Despite numerous works on this issue, SJPWs with
frequencies higher than $\omega_J$ were not discussed before. Here
we study them and prove that the SJPWs spectrum consists of two
branches. The low-frequency branch was described in detail
in~\cite{mdyam1}. Its spectrum $\omega(q)$ follows the ``vacuum
light line'', $q=\omega/c$, and deviates from it at frequencies
close to $\omega_J$. The new branch of SJPWs predicted here starts
at the frequency $\omega=\omega_1^{\rm vac} $ and follows the
vacuum light line for $\omega \ll \omega_2$. For frequencies of
the order of $\omega_2$, the dispersion curve $\omega (q)$
strongly deviates from the vacuum light line and stops at the
frequency $\omega = \omega_2$ when $q\approx 1/\lambda_{ab}$.

Thus, SJPWs do not exist within the frequency gap
$\omega_J<\omega<\omega_1^{\rm vac} $. On the other hand, as shown
here, this is actually the range where the NRI can be observed
(for waves incident from the vacuum onto the
layered-superconductor boundary). Hence, some kind of
complementarity between the NRI and surface waves is established
in this paper.

\begin{figure}
\includegraphics*[width=15 cm]{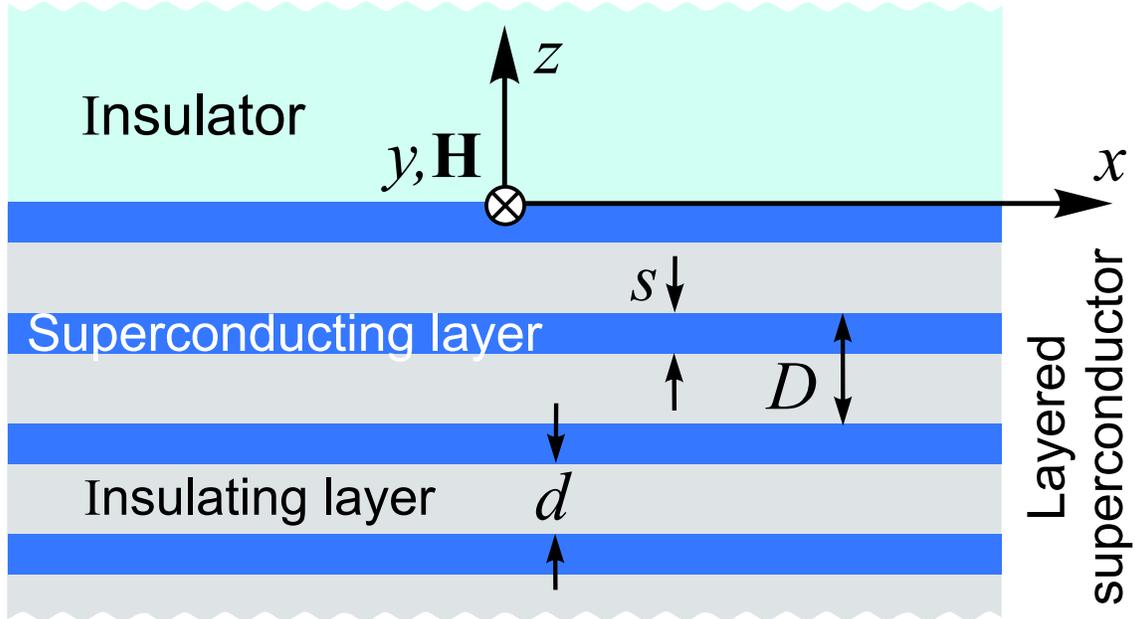}
\caption{(Color online) Geometry for studying waves in layered
superconductors. The interface $z=0$ divides the layered
superconductor from an insulator with dielectric
constant~$\varepsilon_{\rm ext}$.}\label{fig1}
\end{figure}

\textit{Conditions for the observation of NRI}.--- Consider a
layered structure consisting of superconducting and dielectric
layers with thicknesses $s$ and $d$, respectively (see
Fig.~\ref{fig1}). We study the transverse-magnetic JPWs
propagating with wave-vector $\textbf{k}_s=(q,0,\varkappa_s)$ and
having the electric, $\mathbf{E}^s = \{E_x^s, 0, E_z^s\}$, and
magnetic, $\mathbf{H}^s = \{0, H^s, 0\}$, components proportional
to $\exp[i(qx + \varkappa_s z - \omega t)]$. The coordinate system
is shown in Fig.~\ref{fig1}.

The electromagnetic field inside the layered superconductor is
determined by the distribution of the gauge invariant phase
difference $\varphi(x,z,t)$ of the order parameter between
neighboring layers. This phase difference can be described by a
set of coupled sine-Gordon equations (see
review~\cite{Thz-rev-2008-July}). In the continuum and linear
approximation, $\varphi(x,z,t)$ can be excluded from the set of
equations for electromagnetic fields, and the electrodynamics of
layered superconductors can be described in terms of an
anisotropic frequency-dependent dielectric permittivity with
components $\varepsilon_{c}(\Omega)$ and
$\varepsilon_{ab}(\Omega)$ across and along the layers,
respectively~\cite{negref}. In Ref.~\cite{helm}, the effect of
spatial dispersion in $\varepsilon_{c}$, related to the capacitive
interlayer coupling, was taken into account. Here we do not
consider this effect because it is only important for a very
narrow frequency range near $\omega_J$~\cite{helm}.

In the limit $s/d \ll 1$, the equations for
$\varepsilon_{c}(\Omega)$ and $\varepsilon_{ab}(\Omega)$ can be
written as
\begin{equation}\label{epsilon2}
\begin{array}{l}
\varepsilon_{c}(\Omega)=\varepsilon\left(1-\ds\frac{1}{\Omega^2}+i\nu_{c}
\ds\frac{1}{\Omega}\right),\vspace{0.2cm} \\
\varepsilon_{ab}(\Omega)=\varepsilon\left(1-\ds\frac{1}{\Omega^2}
\gamma^2+i\nu_{ab}\ds\frac{1}{\Omega}\gamma^2\right).
 \end{array}
\end{equation}
Here we introduce the dimensionless parameters
$\Omega=\omega/\omega_J$, $\nu_{ab}= 4\pi
\sigma_{ab}/\varepsilon\omega_J\gamma^2$, and $\nu_{c}= 4\pi
\sigma_{c}/\varepsilon\omega_J$. The relaxation frequencies
$\nu_{ab}$ and $\nu_{c}$ are proportional to the averaged
quasi-particle conductivities $\sigma_{ab}$ (along the layers) and
$\sigma_{c}$ (across the layers), respectively; $\omega_J = (8\pi
e D j_c/\hbar\varepsilon)^{1/2}$ is the Josephson plasma
frequency. The latter is determined by the critical Josephson
current density $j_c$,  the interlayer dielectric constant
$\varepsilon$, and the spatial period of the layered structure
$D=s+d \approx d$.

Analyzing the relations for $\varepsilon_{c}(\Omega)$ and
$\varepsilon_{ab}(\Omega)$, we conclude that their real parts have
different signs in a wide frequency range:
$\omega_J<\omega<\omega_2 = \omega_J \gamma$ (or
$1<\Omega<\Omega_2 = \gamma \gg 1$). In this frequency range, the
$z$-components of the group velocity and the Poynting vector of
the \emph{bulk JPWs} are directed opposite to the $z$-component of
the wave-vector $\textbf{k}_s$, and this, at first sight,
corresponds to having a NRI. However, a more careful analysis
shows that this can only be observed for a much narrower frequency
interval. To verify this, one can consider the well-known
dispersion relation for the normal component $\varkappa_s$ of the
JPW wave-vector,
\begin{equation}\label{DispRel-Cont_EPS}
\varkappa_s^2 =
\varepsilon_{ab}(\Omega)\left[k^2-\frac{q^2}{\varepsilon_{c}(\Omega)}\right],
\end{equation}
that follows directly from Maxwell's equations. Obviously, the
JPWs can propagate only when ${\rm Re}(\varkappa_s^2)>~0$, where
``${\rm Re}$'' stands for the real part. Neglecting dissipation,
the permittivity $\varepsilon_{ab}$ is negative for the frequency
region $\omega_J<\omega<\omega_2$ considered here. Consequently,
JPWs can propagate if the factor
$\left[k^2-{q^2}/{\varepsilon_{c}(\Omega)}\right]$ in
Eq.~(\ref{DispRel-Cont_EPS}) is negative. Using
Eq.~(\ref{epsilon2}) for $\varepsilon_{c}(\Omega)$, one can
conclude that this factor is negative \textit{only} when
$1<\Omega^2<1+q^2\lambda_c^2$. For a wave incident, at an angle
$\theta$, from the insulator with dielectric constant
$\varepsilon_{\rm ext}$ onto the layered superconductor, we have
$q=(\omega\varepsilon_{\rm ext}^{1/2} /c)\sin\theta$. Thus, a NRI
can be observed for waves with incident angles \emph{higher than
the critical value} $\theta_{\rm crit}$ defined by the equation,
\begin{equation}\label{theta-criti}
\sin(\theta_{\rm crit}) =
\sqrt{\varepsilon_{c}(\Omega)/\varepsilon_{\rm ext}}.
\end{equation}
It is important to note that, due to the negative sign of the
permittivity $\varepsilon_{ab}$ in Eq.~(\ref{DispRel-Cont_EPS}),
the incident wave penetrates the superconductor for $\theta >
\theta_{\rm crit}$, and totally reflects from it for $\theta <
\theta_{\rm crit}$, contrary to the standard case of waves
incident onto the interface dividing two usual right-handed media.

For $\varepsilon_{\rm ext} < \varepsilon$, Eqs.~(\ref{epsilon2}),
(\ref{theta-criti}), and the inequality $\sin(\theta_{\rm crit})
\leq~1$, provide the conditions,
\begin{equation}\label{omega10}
\omega_J<\omega<
\omega_1=\omega_J\left(\frac{\varepsilon}{\varepsilon-\varepsilon_{\rm
ext}}\right)^{1/2}.
\end{equation}
Thus, the NRI for a layered superconductor bounded by an insulator
with $\varepsilon_{\rm ext} < \varepsilon$ can only be observed in
the frequency range $\omega_J<\omega<\omega_1=\omega_J \Omega_1$.
This frequency window can be expanded if one uses an insulator
with large enough permittivity $\varepsilon_{\rm ext}$. Only for
insulators with very high permittivity $\varepsilon_{\rm ext} >
\varepsilon$, the negative refraction can occur in the whole
frequency interval $\omega_J<\omega<\omega_2$. The frequency range
for the existence of the bulk JPWs in superconductors with
capacitive interlayer coupling was derived in~\cite{helm1}. If the
constant of this coupling tends to zero, the frequency range
obtained in~\cite{helm1} coincides with Eq.~(\ref{omega10}). Below
we consider waves in the insulator-layered superconductor system
with $\varepsilon_{\rm ext} < \varepsilon$ for frequencies
$\omega>\omega_1$.

\textit{Surface Josephson plasma waves above $\omega_1$}.--- When
$\omega_J<\omega<\omega_2$ and the factor in
Eq.~(\ref{DispRel-Cont_EPS}) is positive, the $z$-component
$\varkappa_s$ of the wave-vector $\textbf{k}_s$ becomes imaginary.
This means that the wave damps into the layered superconductor. On
the other hand, for $q>\omega \sqrt{\varepsilon_{\rm ext}}/c$ the
wave damps also into the insulator above the layered
superconductor. These are the characteristic features of the
surface waves discussed in this section.

Consider an interface (the $xy$-plane) separating an insulator
($z>0$ in Fig.~\ref{fig1}) and a layered superconductor
($z\leq0$). We now consider a linear surface transverse-magnetic
monochromatic wave propagating along the $x$-axis (i.e.,
proportional to $\exp[i(qx - \omega t)]$) and \emph{decaying} into
both, the insulator and layered superconductor, away from the
interface $z=0$.

\begin{figure}[hbpt]
\includegraphics[width=15 cm]{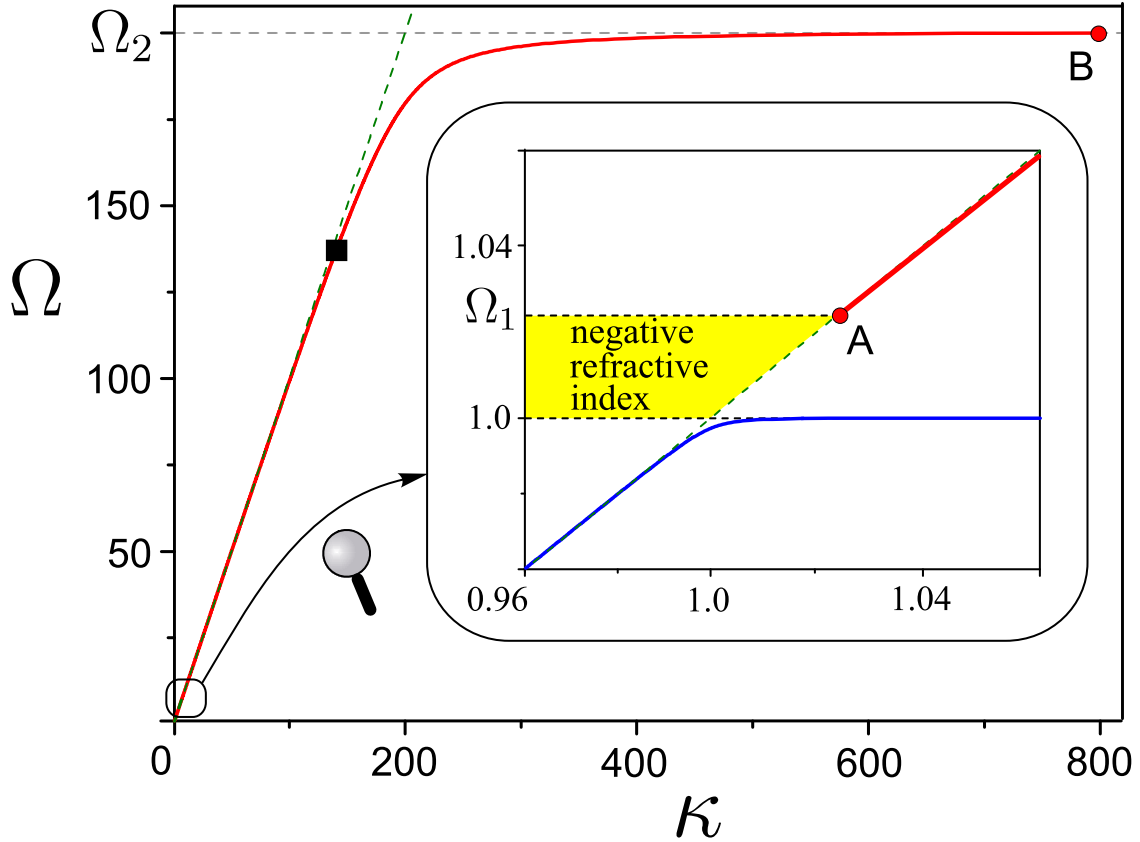}
\caption{(Color online) The dispersion curve
($\Omega=\omega/\omega_J$ versus $\kappa=cq/\omega_J
\varepsilon_{\rm ext}^{1/2}$) for SJPWs at the vacuum-layered
superconductor interface. The values of the parameters are:
$\gamma=200, \, \varepsilon = 16$. Inset: zoom-in of the spectrum
near the point $(\kappa=1, \, \Omega=1)$. Points A and B
correspond to the beginning and end of the high-frequency branch.
The green dashed line is the vacuum light line $\Omega=\kappa$.
}\label{Disp}
\end{figure}

Performing the standard procedure for searching surface waves
(i.e., solving the Maxwell equations for the insulator and layered
superconductor with proper boundary conditions at the interface
between them) we obtain the \emph{dispersion relation} for the
surface Josephson plasma waves:
\begin{equation}\label{Disp-general}
\kappa (\Omega) =
\Omega\left(\varepsilon_c(\Omega)\frac{\varepsilon_{\rm
ext}-\varepsilon_{ab}(\Omega)} {\varepsilon_{\rm
ext}^2-\varepsilon_c(\Omega)\varepsilon_{ab}(\Omega)}\right)^{1/2},
\end{equation}
or, neglecting dissipation,
\begin{equation}\label{Disp-general2}
\kappa (\Omega)=\Omega \left( \frac{\gamma^2-\Omega^2 +\Omega^2
\varepsilon_{\rm ext}/\varepsilon}{\gamma^2
-\Omega^2+\Omega^4\varepsilon_{\rm ext}^2/ (\Omega^2-1)
\varepsilon^2}\right)^{1/2}.
\end{equation}
Here the dimensionless wave-vector is defined as
$\kappa=cq/\omega_J \varepsilon_{\rm ext}^{1/2}$. Equation
(\ref{Disp-general2}) describes two branches of the dispersion
curve for the SJPWs (see Fig.~\ref{Disp}). The first branch exists
in the low frequency range,  $0<\omega<\omega_J$, and it was
studied before in \cite{mdyam1}. The second (predicted here)
branch starts from the light line $\omega = cq/\varepsilon_{\rm
ext}^{1/2}$ (or $\Omega = \kappa$) at $\omega=\omega_1$ (point A
in Fig.~\ref{Disp}), then follows this line, deviates from it at
$\omega \sim \omega_2=\gamma \omega_J$, and stops at the point
where $q = \gamma \omega_J \varepsilon ^{1/2}/c, \, \omega =
\omega_2$ (point B in Fig.~\ref{Disp}).

Thus, there exists a frequency gap,  $\omega_J<\omega<\omega_1$,
in the spectrum of the SJPWs. We emphasize that the NRI should
only exist within this gap.
When the permittivity $\varepsilon_{\rm ext}$ of the insulator
increases, the point A in Fig.~\ref{Disp} moves towards point B,
and the gap in the SJPW spectrum increases. When $\varepsilon_{\rm
ext} = \varepsilon(1 - 1/\gamma^2)  \approx \varepsilon$, the
points A and B coincide, and the high-frequency branch in the
SJPWs spectrum disappears.

Note that the Josephson current is small with respect to the
displacement current at high frequencies, $\Omega \gg 1$. In this
case, we can omit 1 in the denominator in
Eq.~(\ref{Disp-general2}). This corresponds to the dispersion
relation for a periodic layered structure \emph{without coupling
between superconducting layers}. Specifically, the interlayer
Josephson coupling is responsible for the appearance of a
frequency gap, $\omega_J<\omega<\omega_1$, in the spectrum of
SJPWs.

\begin{figure}\centering
\includegraphics[width=15 cm]{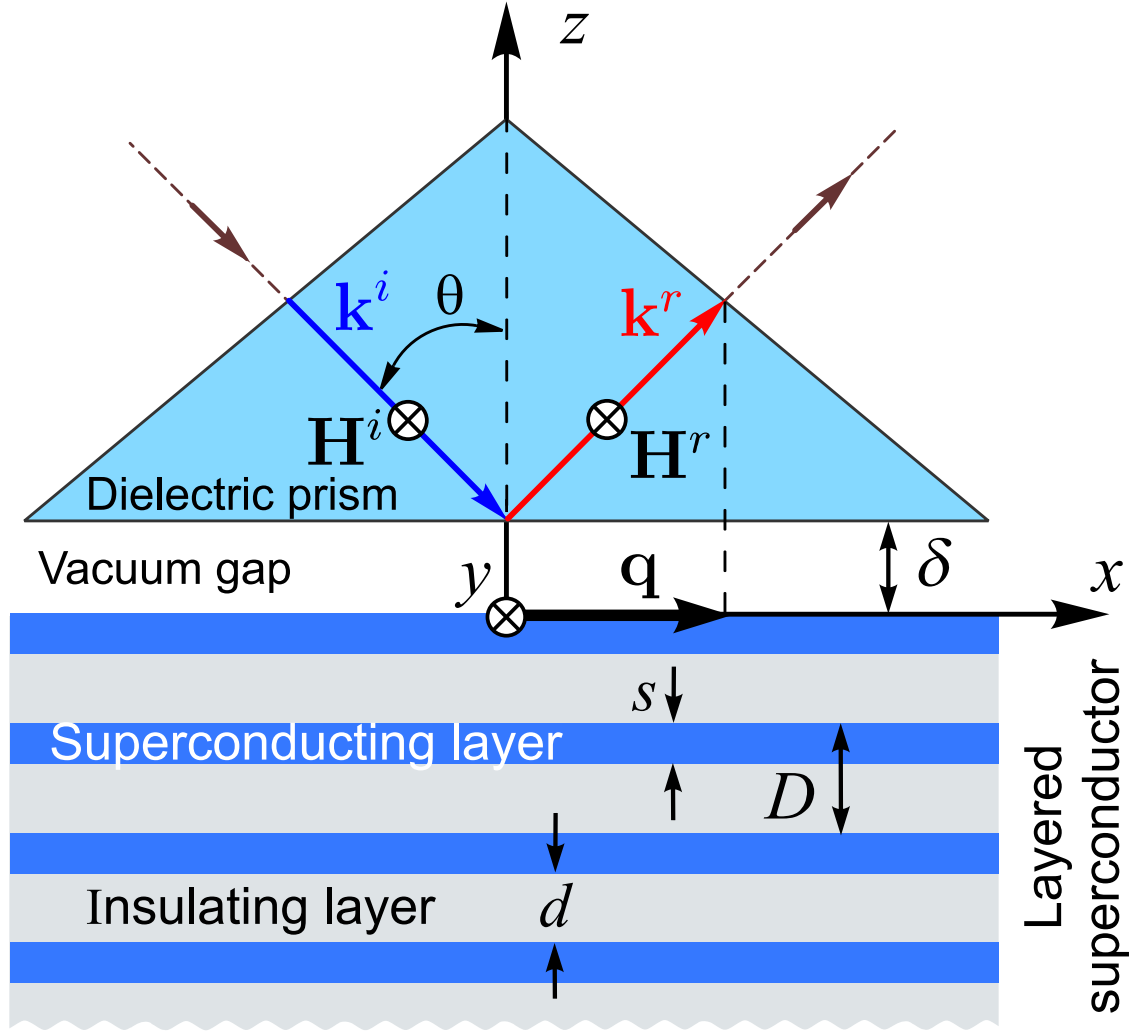}
\caption{(Color online) A dielectric prism with permittivity
$\varepsilon_p$ is separated from a layered superconductor by a
vacuum gap of thickness $\delta$. An electromagnetic wave with
incident angle $\theta$, exceeding the limit angle $\theta_t =
\arcsin(\varepsilon_p^{-1/2})$ for total internal reflection, can
excite surface waves that satisfy the following resonance
condition: $\omega \varepsilon_p^{1/2} \sin\theta=cq$. Here ${\bf
k}^i$ and ${\bf k}^r$ are the wave-vectors of the incident and
reflected waves associated with the magnetic field amplitudes
${\bf H}^i$ and ${\bf H}^r$. The resonance excitation of surface
waves by the incident wave produces a strong suppression of the
reflected wave. }\label{Prizm_Geom}
\end{figure}

\textit{Excitation of the SJPWs above $\omega_1$}.--- It is known
that the excitation of surface waves is accompanied by the
so-called Wood anomalies of the reflectivity and transmissivity
coefficients (see, e.g., \cite{Agrmils}). These resonance
anomalies can result in the complete suppression of the
reflectivity by a proper choice of parameters. Here we consider
the excitation of high-frequency SJPWs ($\Omega \gg 1$) by a wave
incident from a dielectric prism with permittivity $\varepsilon_p$
onto a layered superconductor separated from the prism by a vacuum
gap of thickness $\delta$ (the so-called
``attenuated-total-reflection'' method for producing surface
waves, see Fig.~\ref{Prizm_Geom}).

\begin{figure}
\includegraphics[width=15 cm]{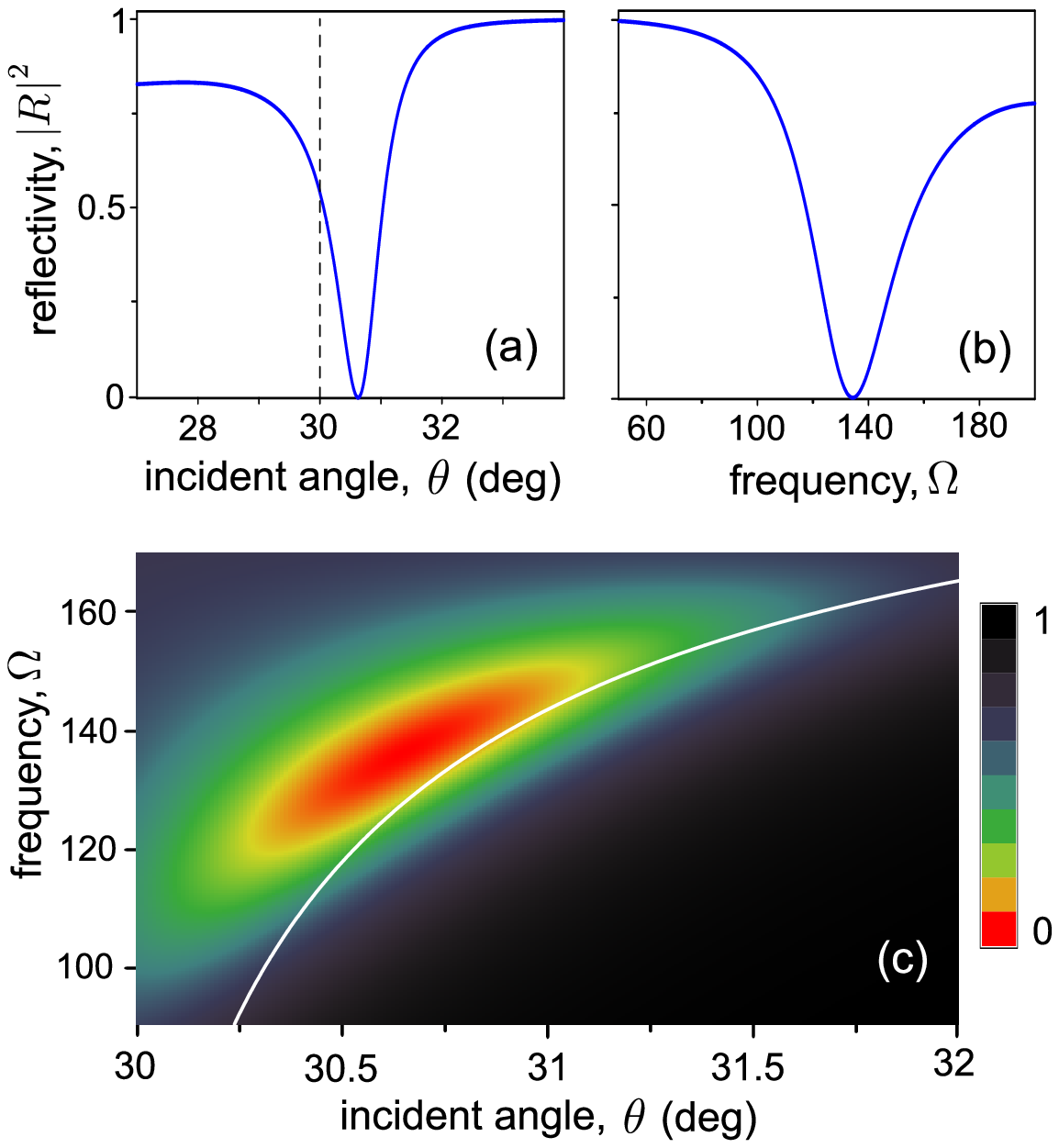}
\caption{(Color online) (a) The reflectivity coefficient $|R|^2$
versus the incident angle $\theta$ calculated numerically for the
parameters $\nu_{ab}=10^{-3}$, $\gamma=200$, $\varepsilon=16$,
$\varepsilon_p=4$, and $\Omega=135$. These parameters correspond
to the solid square on the dispersion curve in Fig.~\ref{Disp}.
The thickness $\delta$ of the vacuum gap (see
Fig.~\ref{Prizm_Geom}) is one wavelength, $ k \delta=2\pi$. The
vertical dashed line at $\theta = 30^\circ$ corresponds to the
limit angle of the total internal reflection. (b)~The reflectivity
coefficient $|R|^2$ versus frequency, $\Omega=\omega/\omega_J$,
for $\theta =30.6^\circ$. (c)~Color contour plot of the
reflectivity coefficient $|R|^2$ in the plane $(\theta,\,
\Omega)$. The dispersion relation for the waves in the
dielectric-vacuum-layered superconductor system in
Fig.~\ref{Prizm_Geom} is presented by the solid white
curve.}\label{R_abc}
\end{figure}

The suppression of the specular reflectivity $|R|^2=|H^r/H^i|^2$
due to the resonant excitation of the surface waves can be
observed by changing the incident angle at a given frequency or by
changing the frequency at a given incident angle, as demonstrated
in Fig.~\ref{R_abc} (a, b). Here $H^i$ and $H^r$ are the magnetic
field amplitudes of the incident and reflected waves,
respectively. Figure \ref{R_abc} (c) shows the sharp decrease of
the reflectivity in the $(\theta, \Omega)$ plane.

\textit{Conclusions}.--- Here we predict the existence of a new
branch of SJPWs in layered superconductors for the frequency range
higher than the Josephson plasma frequency, which is a very
unusual phenomenon for plasma-like media. It is important that a
NRI can only be observed for frequencies within the gap in the
spectrum of the SJPWs. Thus, some kind of complementarity between
NRI and surface waves is established in this paper. We have also
described the excitation of these SJPWs by means of the
attenuated-total-reflection method. 

We gratefully acknowledge
partial support from the NSA, LPS, ARO, and NSF grant
No.~0726909.

\end{document}